\documentclass{article}
\usepackage{amssymb,amsthm,amsmath}
\usepackage{amsfonts}
\usepackage{graphics}
\usepackage[latin1]{inputenc}
\usepackage{color}
\usepackage{float}
\usepackage{lmodern} %pour avoir une police de meilleur qualite
\usepackage[left=1.5in,right=1.3in,top=1.1in,bottom=1.1in,includefoot,includehead,headheight=13.6pt]{geometry}

\usepackage[dvips]{graphicx,epsfig}
\usepackage{subfigure}
\usepackage{latexsym,euscript,epic,eepic}
\usepackage{float,caption}
\usepackage{subfigure}
\usepackage{bbding}
\newtheorem{Theo}{Theorem}

\newtheorem{Lem}{Lemma}

\begin{document}
\title{{\sc Modeling Risk via Realized HYGARCH Model
}}
\author{El Hadji Mamadou Sall$^{(1,*)}$,
El Hadji Deme$^{(1)}$ and
Abdou Ka Diongue $^{(1)}$\\
~~\\
\small $^{(1)}$ \textsl{LERSTAD, UFR SAT Universit\'e Gaston Berger, BP 234 Saint-Louis, S\'en\'egal}~~~~~~~~~~~~~~~~~~~~~~~~~~~~\\
$^{(*)}$ Corresponding autor: \textit{elhadjimadou.sall@yahoo.fr}
~~~~~~~~~~~~~~~~~~~~~~~~~~~~~\\
}
 \date{
 \;
 }
\maketitle
\noindent\rule[2pt]{\textwidth}{0.5pt}
\textbf{\large{Abstract.}} ~~\\
In this paper, we propose the realized Hyperbolic GARCH model for the joint-dynamics of low-frequency returns and realized measures that generalizes the realized GARCH model of Hansen et \textit{al.} in \cite{hansen2012realized} as well as the FLoGARCH model introduced by Vander Elst in \cite{vander2015fir}. This model is sufficiently flexible to capture both long memory and asymmetries related to leverage effects. In addition, we will study the strictly and weak stationarity conditions of the model. To evaluate its performance, experimental simulations, using the Monte Carlo method, are made to forecast the Value at Risk (VaR) and the Expected Shortfall (ES). These simulation studies show that for ES and VaR forecasting, the realized Hyperbolic GARCH (RHYGARCH-GG) model with Gaussian-Gaussian errors provide more adequate estimates than the realized Hyperbolic GARCH model with student-Gaussian errors.\\
\noindent\rule[2pt]{\textwidth}{0.5pt}
\section{Introduction}
Volatility forecast of asset returns is very important for option pricing as well as risk management. Since the Autoregressive Conditional Heteroskedasticity model (ARCH) introduced by Engle \cite{engle1982autoregressive} and generalized by Bollerslev \cite{bollerslev1986generalized} are widely used to study the properties of volatility for economic and financial data. However, there are several shortcomings with using GARCH model for risk management or forecasting volatility. The major issue is the persistence of variance that evolves through time which the GARCH model fails to address. To overcome this problem, many models are introduced in the literature.   Among others, we can cite the IGARCH model of Engle and Bollerslev in \cite{engle1986modelling}, the  FIGARCH model of  Bollerslev and Baillie in \cite{baillie1996fractionally}, the FIEGARCH model of Bollerslev and Baillie in \cite{lopes2014theoretical}, the HYGARCH model of Davidson in \cite{davidson2004moment} where the conditional variance is a convex combination of the conditional variances of GARCH and FIGARCH models, the new HYGARCH of Li et \textit{al. } in \cite{li2015new}, and the S-HYGARCH model of Diongue and Guegan in \cite{diongue2007stationary}. However, as stated by Hansen et \textit{al.} in \cite{hansen2012realized} and discussed by Andersen et \textit{al.} in \cite{andersen2003modeling},  a single return only offers a weak signal about the current level of volatility. Therefore, the implication is that GARCH models are poorly suited for situations where volatility changes rapidly to a new level. Indeed, GARCH models are slow at catching up, and it will take many periods for the conditional variance to reach its new level. To alleviate this problem, researchers proposed to incorporate realized measures in the GARCH model.\\

\noindent Moreover, with the advent of high-frequency data, several measures have been developing in the literature, such as the Realized Variance and Realized Kernel, among many others Anderson and Bollerslev, Barndorff-Nielsen and Shephard, and Barndorff-Nielsen et \textit{al.} in \cite{andersen1998answering, barndorff2002econometric, barndorff2008designing}. All of these measures provide more information on the current level of volatility compare to the square of returns. This aspect makes that realized measures have attracted recently the attentions of financial econometricians as an accurate estimator of volatility. For instance, Engle in \cite{engle2002new}  introduced the GARCH-X model by including realized measures in the GARCH equation. In \cite{hansen2012realized}, Hansen et \textit{al.} proposed the Realized GARCH model by completing GARCH-X models with a measurement equation for the realized measure. Later, in \cite{hansen2016exponential}, Hansen and Huang introduced the Realized EGARCH to capture the asymmetries related to leverage effects while Takahashi et \textit{al.} \cite{takahashi2009estimating} have extended the stochastic volatility model in the same direction. Watanabe in \cite{watanabe2012quantile} used daily returns, realized volatility and realized kernel of the S\&P 500 stock index to quantile forecasts and found that realized GARCH model with skewed Student's -t distribution performs better than that with normal and Student's t-distributions. In addition,  Vander Elst \cite{vander2015fir} proposed FLoGARCH models (FractionaLly integrated realized vOlatility GARCH)  to capture also the property of long memory observed on the realized measure. He showed that, using the S\&P 500 daily return,  FLoGARCH models provide more accurate forecasts than realized GARCH models and FIGARCH models. However, FLoGARCH models used FIGARCH models, for which the existence of a stationary solution with infinite variance was not yet proved (see, Giraitis, Leipus and Surgailis in \cite{giraitis2009arch}, Tayefi and Ramanathan in \cite{tayefi2012overview} among others), as GARCH equation.\\

\noindent To overcome the problem of infinite variance in FIGARCH models, in this work, we focus on another class of asymmetric long memory GARCH process that belong to the family of realized GARCH models introduced by Hansen et \textit{al}. In particular, we introduce the realized Hyperbolic GARCH model (RHYGARCH) which extended the FLoGARCH model of Vander Elst \cite{vander2015fir}).  Conditions for strict and weak stationary of the model will be established. In addition, by considering two types of model (the realized HYGARCH model with Gaussian Gaussian error and the realized HYGARCH model with student-t Gaussian error), experimental simulations, using Monte Carlo method, are implemented for quantile forecasts (Value at Risk (VaR) and Expected Shortfall (ES)). For empirical studies, the model can be applied to the S\&P 500 (SPY) stock index as in Vander Elst \cite{vander2015fir}).

%\noindent The main objective of the 1998 Basel Accord is to develop a risk-based capital framework that strengthens and stabilizes the banking system. In 1996, this accord was revised to take into account the importance of market risk. Capital requirements are a common aspect of the regulation of financial institutions. They were implemented with a list of standardized rules that appear simple and robust but have the drawback to not be sensitive enough to the risk profile of the institutions. 

\noindent This paper is organized as follows. In Section 2, we present the realized HYGARCH model. In Section 3, we study the stationarity conditions, while in Section 4, we present the likelihood estimation functions.  In Section 5, we present the forecasting method used to estimate the VaR and the ES. Section 6 investigates Monte Carlo simulation experiments in order to evaluate the finite sample properties of the model. Finally, Section 7 concludes.

\section{The model}
\subsection{General structure of realized HYGARCH model}
The general formula of the realized GARCH model is given by:
\begin{align}
      &r_t=h_t^{1/2}z_t, \label{eq1} \\ 
     & h_t=u\left(h_{t-1},\cdots,h_{t-p},x_{t-1},\cdots,x_{t-q}\right),\label{eq2}\\ 
    & x_t=m\left(h_t,z_t,u_t\right),  \label{eq3}
\end{align}
\noindent where $r_t$ is the return, $x_t$ a realized measure of volatility, $\left(z_t\right)_t$ are identically independently distributed (i.i.d) with  mean zero  and variance one, $\left(u_t\right)_t$ are also i.i.d  with mean zero and variance $\sigma_u^{2}$. Here $\left(z_t\right)_t$ and $\left(u_t\right)_t$  are mutually independent. In addition, $E\left(r_t\mid \mathcal{F}_{t-1}\right)=0$ and $E\left(r_t^2\mid \mathcal{F}_{t-1}\right)=h_t$,  where $\mathcal{F}_{t-1}$ denotes the sigma field generated by the past information up to $t-1$. More specifically $\mathcal{F}_t=\sigma\left(X_t,X_{t-1},\cdots\right)$, with $X_t=\left(r_t,x_t\right)$. We label equation (\ref{eq1}) as return equation, equation (\ref{eq2}) as the GARCH equation and equation (\ref{eq3}) as the measurement equation. \\

\noindent Most (if not all) variants of ARCH and GARCH models are nested in the Realized HYGARCH framework. The nesting can be achieved by setting $x_t = r_t$ or $x_t=r_t^2$, and the measurement equation is redundant for such models, because it is reduced to a simple identity. However, see Bollerslev in \cite{bollerslev2010volatility}, the interesting case is when $x_t$ is a high-frequency-based realized measure, or a vector containing several realized measures. Next we consider a particular variant of the Realized GARCH model, where an HYGARCH model is considered as GARCH equation.

 \subsection{Realized HYGARCH model}
 
% The use of the model with log-linear specification is justified by the following reasons :
% \begin{itemize}
% \item firstly, an attractive feature of the log-linear realized GARCH model is that it keeps the ARMA structure that characterizes some of the standard GARCH model;
% \item secondly, the log specification ensures a positive variance.
% \end{itemize}
Recall that the log-linear realized GARCH$\left(p,q\right)$ by Hansen et \textit{al.} in \cite{hansen2012realized}  is defined as follows :
 \begin{align}
 & r_t=h_t^{1/2}z_t,  \label{eq4} \\
& \log\left(h_t\right)=\omega^{\prime}+\beta\left(L\right)\log\left(h_t\right)+\alpha\left(L\right)\log\left(x_t\right),  \label{eq5}\\
 & \log\left(x_t\right)=\xi +\phi\log\left( h_t\right) + \tau\left(z_t\right)+u_t, \label{eq6}
 \end{align}
\noindent where $L$ denotes the lag or backshift operator, $\beta\left(L\right)=\beta_1L+\beta_2L^2+\cdots+\beta_pL^p$ and $\alpha\left(L\right)=1+\alpha_1L+\alpha_2\L^2+\cdots+\alpha_qL^q$. The polynomial $\tau\left(z\right)=\tau_1 z+\tau_2\left(z^2-1\right)$ is called leverage function and  facilitate a modeling of the dependence between return shocks and volatility shocks.\\
 
% \noindent All realized HYGARCH model keep the equation of return and equation of measure. The main input is provided in the GARCH equation.
\noindent Remark that the GARCH$(p,q)$ process may  be expressed as an ARMA$\left(m,p\right)$ process :
\[
\left[1-\beta\left(L\right)-\alpha\left(L\right)\right]\log x_t=\omega^{\prime}+\left[1-\beta\left(L\right)\right]v_t,
\]
\noindent where $m=\max\left(p,q\right)$ and $v_t=\log\left(x_t\right)-\log\left(h_t\right)$. Thus, the process $\left(v_t\right)_t$  is interpreted as  "innovations" for the conditional variance. When the polynomial $1-\beta\left(L\right)-\alpha\left(L\right)$ contains a unit root then $\log\left(x_t\right)$ can be defined as an $I\left(1\right)$ process which is written as
\[
\phi\left(L\right)\left(1-L\right)\log\left(x_t\right)=\omega^{\prime}+\left[1-\beta\left(L\right)\right]v_t,
\]
where the polynomial $\phi\left(L\right)=\left[1-\beta\left(L\right)-\alpha\left(L\right)\right]\left(1-L\right)^{-1}$ is of order $m-1$. Letting $\gamma\left(L\right)=1-\beta\left(L\right)-\alpha\left(L\right)$, the model can be rearranged as:
 \begin{eqnarray*}
 \log\left(h_t\right)&=&\omega+\left(1-\frac{\gamma\left(L\right)}{1-\beta\left(L\right)}\right)\log\left(x_t\right){}
 \nonumber\\
 &=&\omega+\pi\left(L\right)\log\left(x_t\right)  
 \end{eqnarray*}
 \noindent where $\pi\left(L\right)=1-\frac{\gamma\left(L\right)}{1-\beta\left(L\right)}$,  $\omega=\frac{\omega^{\prime}}{1-\beta\left(1\right)}$ . Notice that \noindent Vander Elst \cite{vander2015fir} replaces this expression with a fractional difference given by
\[
  \pi\left(L\right)=1-\frac{\gamma\left(L\right)\left(1-L\right)^d}{1-\beta\left(L\right)},
\]
where $\gamma\left(z\right)=0$ has roots outside the unit circle, allows for long-range dependencies in $\log\left(x_t\right)$. The model can be then written as :
\begin{equation}\label{eq7}
\log\left(h_t\right)=\omega+\left\{1-\gamma\left(L\right)\left[1-\beta\left(L\right)\right]^{-1}\left(1-L\right)^d\right\}\log\left(x_t\right).  \end{equation}
\noindent Equation (\ref{eq7}) is considered in the FLoGARCH model, introduced by Vander Elst \cite{vander2015fir}, as the volatility equation. However, the FIGARCH$\left(p, d, q\right)$ model capture long-range dependence that possesses hyperbolic decay of ACF but has infinite variance which limits its application. Therefore, for the sake of generality, we write the volatility process as :
 \begin{equation}\label{eq8}
 \log\left(h_t\right)=\omega+\delta\left\{1-\gamma\left(L\right)\left[1-\beta\left(L\right)\right]^{-1}\left(1-L\right)^d\right\}\log\left(x_t\right),
 \end{equation}
\noindent with $d\geqslant 0$ and  $ 0\leqslant\delta\leqslant 1$. Equation (\ref{eq8}) which is the  GARCH equation for the realized HYGARCH$\left(p,d,q\right)$ model can be viewed as the HYGARCH$\left(p,d,q\right)$ model of Li et \textit{al.} in  \cite{li2015new} that  has  a  form  nearly the FIGARCH process while allowing the existence of finite variance. This model contains several other extensions among others, the FLoGARCH model when $\delta=1$ and the realized GARCH model if $d=1$ and $\delta\leq 1$.
Equations (\ref{eq4}),(\ref{eq6}) and (\ref{eq8}) define the realized HYGARCH model.\\

\noindent For the rest of the paper, the realized HYGARCH $\left(1,d,1\right)$ model  is considered. It can be written as:
 \begin{align}   
&r_t=h_t^{\frac{1}{2}}z_t \\
&\log h_t=\omega +\delta\left[1-\frac{1-\gamma L}{1-\beta L}\left(1-L\right)^d\right]\log x_t \\
&\log x_t=\epsilon+\phi \log h_t +\tau_1z_t+\tau_2\left(z_t^2-1\right)+\theta_u u_t, 
\end{align}
\noindent with $\left(1-L\right)^d=\displaystyle\sum_{k=0}^{\infty}\frac{\Gamma\left(d+1\right)\left(-L\right)^k}{\Gamma\left(k+1\right)\Gamma\left(d-k+1\right)}$,\\  $r_t$ is the percentage log-return for day $t$, $z_t\overset{iid}{\sim}D_1\left(0,1\right)$, $u_t\overset{iid}{\sim}D_2\left(0,\sigma_u^2\right)$ and $x_t$ is the realized measure.\\

\noindent  Hansen et \textit{al.} in \cite{hansen2012realized} used the RV (Realized Variance) and RK (Realized Kernel) as realized measures $x_t$ in the realized GARCH model and considered Gaussian distributed errors $D_1\left(0,1\right)=N\left(0,1\right)$ and $D_2\left(0,\sigma_u^2\right)=N\left(0,\sigma_u^2\right)$.  Gerlach and Chaowang in \cite{gerlach2016forecasting} used standardized student-t as $D_1\left(0,1\right)$ while Contino and Gerlach in \cite{contino2017bayesian} and Watanabe in \cite{watanabe2012quantile} considered a Skew student-t as $D_1\left(0,1\right)$.   

\section{Existence of the second-order stationary solution}
 
\noindent In this section, we study the existence conditions of stationary solution to equations (\ref{eq4}), (\ref{eq6}) and (\ref{eq8}). More precisely, we investigate the strict and weak stationary conditions of the random sequence $\left(\log h_t, t \in \mathbb{Z}\right)$. Thus, let $\tilde{h}_t=\log h_t$ and $\tilde{x}_t=\log x_t$ be related as
\begin{align}
\label{eq9}
 \tilde{h}_t=&\omega+\psi\left(L\right)\tilde{x}_t\\
 =&\omega+\sum_{i=1}^{\infty}\psi_i\tilde{x}_{t-i},
\end{align}
with 
\begin{align} 
\label{eq10}
\psi\left(L\right)=\delta\left\{1-\phi\left(L\right)\left[1-\alpha\left(L\right)\right]^{-1}\left(1-L\right)^d\right\}.
\end{align}
\begin{Lem}\label{le1} 
Let $\left(\tilde{h}_t\right)$ be the process defined by (\ref{eq8}), if the conditions 
\begin{equation}\label{eq11}
\vert \phi\sum_{i=1}^{\infty} \psi_i \vert <1\ \ and\ \ \sum_{i=1}^{\infty}\vert \psi_i\vert <\infty
\end{equation}
are satisfied then
\begin{equation}\label{eq12}
\tilde{h}_t=\sum_{l=0}^{\infty} H_l(t),
\end{equation}

where
\begin{eqnarray*}
 H_l\left(t\right)&=&\sum_{i_1,i_2,\cdots,i_l=1}^{\infty}\phi^{l-1} \psi_{i_1}\psi_{i_2}\cdots, \psi_{i_l}\left(\omega\phi+v_{t-i_1-i_2-\cdots -i_l}\right) \ for\ \  l\geq 1; \\
 H_0\left(t\right)&=&\omega.  
\end{eqnarray*}

\end{Lem}

\noindent We resume in the following theorem, the necessary and sufficient conditions for the existence of a stationary solution for the process (\ref{eq4}), (\ref{eq6}) and (\ref{eq8}).
\begin{Theo}\label{theo1}
Let $\tilde{h}_{t\in\mathbb{Z}}$ be the process  defined by (\ref{eq8}):
\begin{enumerate}
\item If the conditions
\begin{eqnarray}\label{eq13}
\phi\sum_{i=1}^{\infty}\psi_i<1\ \ and \ \  \sum_{i=1}^{\infty}\vert\psi_i\vert<\infty
\end{eqnarray}
are satisfied, where the weights $\left(\psi_i\right)_{i\in \mathbb{N}}$ are given in (\ref{eq10}), then by restricting,  the first moment of the bivariate process $\left(\tilde{h}_t, \tilde{x}_t\right)$ exists and is given by :
\[
E\left(\tilde{h}_t\right)=\frac{\omega+\xi\sum_{i=1}^{\infty}\psi_i}{1-\phi\sum_{i=1}^{\infty}\psi_i}=\frac{\omega+\xi\psi(1)}{1-\phi\psi(1)}
\] and 
\[
E\left( \tilde{x}_t\right)=\frac{\xi+\phi\omega}{1-\phi\psi(1)}.
\]
\item  If the conditions 
\begin{equation}\label{eq14}
E\left(z_0^3\right)<\infty,\quad E\left(z_0^4\right)<\infty,\quad \omega=0,\quad \phi>0,\quad \psi_i\geq 0\quad \forall\quad i\geq 1,   
\end{equation}
and Lemma \ref{le1} are satisfied, then the second moment of the process $\tilde{h}_t$ exists.
\end{enumerate}
\end{Theo}
\noindent Under the conditions provided in Theorem \ref{theo1}, we investigate the strict and weak stationary solution for the realized HYGARCH model. The results concerning $\tilde{h}_t$ are resumed in the following theorem.

\begin{Theo}\label{theo2}
Let $(\tilde{h}_t)_{t\in\mathbb{Z}}$ be the process defined by (\ref{eq8})
\begin{enumerate}
\item Under conditions (\ref{eq13}) and Lemma \ref{le1}, (\ref{eq12}) represents a unique strictly stationary solution for the process $(\tilde{h}_t)_{t\in\mathbb{Z}}$;
\item If, in addition, the condition (\ref{eq14}) is verified then (\ref{eq12}) is also a unique weakly stationary solution.
\end{enumerate}
\end{Theo}

\section{ Likelihood Estimation}
\noindent In this section, we tackle the problem of estimating the parameters of realized HYGARCH$\left(p,d,q\right)$ model. For this, the quasi-maximum likelihood method is used.
\begin{itemize}
\item Following Harry Vander Elst \cite{vander2015fir} where $D_1=N\left(0,1\right)$ and $D_2=N\left(0,\sigma_u^2\right)$, the log-likelihood function for the model is given by :
 \begin{equation}
l\left(r,x,\theta_G\right)=\underbrace{-\frac{1}{2}\sum_{t=1}^n\left(\log 2\pi+\log h_t+\frac{r_t^2}{h_t}\right)}_{=l\left(r\mid x;\theta_G\right)}\underbrace{-\frac{1}{2}\sum_{t=1}^n\left[\log 2\pi+\log\left(\sigma_u^2\right)+\frac{u_t^2}{\sigma_u^2}\right]}_{=l\left(x;\theta_G\right)},
\end{equation}
\noindent  where $u_t=\log x_t-\epsilon-\phi \log h_t-\tau_1z_t-\tau_2\left(z_t^2-1\right)$. This model is denoted by RHYGARCH-GG (RHYGARCH with Gaussian-Gaussian error). The parameters $\pmb{\mathbb{\theta}_G}=\left(w,d,\delta,\pmb{\mathbb{\alpha}},\pmb{\mathbb{\beta}},\epsilon,\phi,\tau_1,\tau_2,\sigma_u\right)$ with $\pmb{\mathbb{\alpha}}=\left(\alpha_1,\cdots,\alpha_q\right)$ and $\pmb{\mathbb{\beta}}=\left(\beta_1,\cdots,\beta_p\right)$.
 \item Under the choice $D_1=t^\ast\left(0,1,\nu\right)$ and $D_2=N\left(0,\sigma_u^2\right)$ as in Gerlach and Chaowang \cite{gerlach2016forecasting} and Contino and Gerlach \cite{contino2017bayesian}, the log-likelihood function for this model is given by:
 \begin{eqnarray} 
l\left(r,x;\theta_t\right)&=&\underbrace{-\frac{1}{2}\sum_{t=1}^n\left[\log  2\pi+\log\left(\sigma_u^2\right)+\frac{u_t^2}{\sigma_u^2}\right]}_{=l\left(x;\theta_t\right)}\nonumber \\
&&
\underbrace{-\sum_{t=1}^n  \left\{A\left(\nu\right)+\log\left[\pi\left(\nu-2\right)\right]+\frac{1}{2}\log\left(h_t\right)+\dfrac{\nu+1}{2}\log\left[1+\frac{r_t^2}{h_t\left(\nu-2\right)}\right]\right\}}_{l\left(r\mid x;\theta_t\right)},
\end{eqnarray}\\
\noindent where $u_t=\log x_t-\epsilon-\phi \log h_t-\tau_1z_t-\tau_2\left(z_t^2-1\right) $ and $t^\ast\left(0,1,\nu\right)=t\left(0,1,\nu\right)\sqrt{\frac{\nu-2}{\nu}} $,which is a student-t distribution with $\nu$ degrees of freedom, scaled to have variance 1, and $A\left(\nu\right)=\log\left[\Gamma\left(\frac{\nu}{2}\right)\right]-\log\left[\Gamma\left(\frac{\nu+1}{2}\right)\right]$. The parameters $\pmb{\mathbb{\theta}_t}$ of this model, denoted by RHYGARCH-tG, is defined as $\pmb{\mathbb{\theta}_t}=\left(w,d,\delta,\pmb{\mathbb{\alpha}},\pmb{\mathbb{\beta}},\epsilon,\phi,\tau_1,\tau_2,\sigma_u,\nu\right)$ with $\pmb{\mathbb{\alpha}}=\left(\alpha_1,\cdots,\alpha_q\right)$ and $\pmb{\mathbb{\beta}}=\left(\beta_1,\cdots,\beta_p\right)$.
 \end{itemize}
 
\section{Forecasting Method}

\subsection{Value-at-Risk Forecasts}

\noindent In order to forecast tail risk in a parametric realized HYGARCH setting, the model is used to estimate a one-ahead volatility forecast at both the 95\% and 99\% confidence level $\alpha$. The conditional one-period-ahead VaR forecast is defined as :
 \[
 \alpha=P\left(r_{t+1}<VaR_{\alpha}\mid\mathcal{F}_t\right),
 \]
\noindent where $r_{t+1}$ is the one-period return from time $t$ to time $t+1$, $\alpha$ is the quantile level and $F_t$ is the informative set at time $t$. For a normal distribution, VaR is calculated via the inverse standard Gaussian CDF, denoted by $\Phi^{-1}\left(\alpha\right)$:
 \[
   VaR_{\alpha}=\sqrt{h_{t+1}}\Phi^{-1}\left(\alpha\right),
   \]
\noindent where $\Phi^{-1}\left(\alpha\right)$ is the inverse standard Gaussian. And similarly for a student-t distribution :
 \[
 VaR_{\alpha}=\sqrt{h_{t+1}}T_{\nu}^{-1}\left(\alpha\right),
  \] 
\noindent where $T_{\nu}^{-1}\left(\alpha\right)$ is the inverse standardized student-t CDF.
 
\subsection{Conditional value-at-risk forecasts}

\noindent The Conditional Value-at-Risk forecasts (CVaR) or Expected Shortfall (ES) is used to estimate a one-ahead volatility forecast, as it has become preferred to the VaR due to the latter's shortcomings. The Expected Shortfall is defined as :
\[   
 ES_{\alpha}=E\left(r_{t+1}\mid r_{t+1}\geq VaR_{\alpha},\mathcal{F}_t \right).
 \] 
\noindent For a normal distribution, the Expected shortfall is calculated via that expression :
\[
ES_{\alpha}=\sqrt{h_{t+1}}\frac{\phi\left(\Phi^{-1}\left(\alpha\right)\right)}{1-\alpha},
\]
\noindent where $\phi\left(x\right)$ and $\Phi^{-1}\left(\alpha\right)$ are the normal probability density function and inverse distribution function respectively. For a student-t distribution, we can derive the Expected shortfall :
\[
ES_{\alpha}=\sqrt{h_{t+1}}\frac{t_{\nu\left[T_{\nu}^{-1}\left(\alpha\right)\right]}}{1-\alpha}\left[\frac{\nu+\left(T_{\nu}^{-1}\left(\alpha\right)\right)^2}{\nu -1}\right],
\]
\noindent where $\nu$ is the estimated degrees of freedom, $t_{\nu}\left(x\right)$ and $T_{\nu}^{-1}\left(\alpha\right)$ are the student-t probability density and inverse cumulative distribution function. 

\section{Simulation study}

\noindent In this section, we have designed and executed Monte Carlo simulation with the aim of analyzing the sampling properties of the MLE estimators for the realized HYGARCH$\left(1,d,1\right)$ model with Gaussian Gaussian error and with student-t Gaussian error. Accross M=1000 Monte Carlo replications and sample size T=1000 and T=3000, two specific models are considered:

\begin{enumerate}
 \item Model 1 
 \[
 r_t=\sqrt{h_t}{z_t},\quad z_t\sim N\left(0,1\right),
 \]
 \[
\log h_t=0.1+0.4\left[1-\frac{1-0.1L}{1-0.4L}\left(1-L\right)^{0.4}\right]\log x_t,
\]
  \[  
   \log x_t=-0.1+1\log h_t-0.08z_t+0.06\left(z_t^2-1\right)+u_t,\quad u_t\sim N\left(0,0.4\right).
     \]
 \item Model 2 
 \[
 r_t=\sqrt{h_t}{z_t},\quad z_t\sim t_3^{\ast}\left(0,1\right),
 \]
  \[
  \log h_t=0.1+0.4\left[1-\frac{1-0.1L}{1-0.4L}\left(1-L\right)^{0.4}\right]\log x_t,
  \]
   \[
   \log x_t=-0.1+1\log h_t-0.08z_t+0.06\left(z_t^2-1\right)+u_t,\quad  u_t\sim N\left(0,0.4\right).
   \]
 \end{enumerate}
\noindent Where $r_t$ is the daily log-return, $x_t$ the daily realized measure and $t^{\ast}$ represents the Student-t distribution standardized to have variance 1. Notice that the chosen  parameters verify the stationary conditions. For other parameters choice, one can refer to Contino and Gerlach in \cite{contino2017bayesian}, and Li et \textit{al.} in \cite{li2015new} for the parameter $\nu$. 
\begin{table}[t]
\caption{Summary statistics for the estimator of the RHYGARCH-GG model, data simulated from Model 1 }
\label{tab1}
\begin{tabular}{@{}llllllll}
\hline
RHGARCH$\left(1,d,1\right)$ &   & T=1000 &  & &  T=3000 &  &\\
parameter & True   &    MSE   &   Mean& True  & MSE &  Mean & \\
\hline
$\omega$ & 0.1 &  0.0030  &0.1178 &0.1 &0.0010 & 0.1062     &    \\
$\gamma$ & 0.1     & 0.0334 & 0.1268 &  0.1   &0.0038   &0.0887 &   \\
$\beta$  & 0.4   &   0.0327& 0.3710& 0.4   & 0.0122  & 0.3683   &   \\
$\delta$ & 0.4 & 0.0546       &  0.4113 & 0.4  &0.0207    & 0.4218 &   \\
$d$      & 0.4 &   0.0573  & 0.4511 & 0.4 &0.0150 & 0.39809 &   \\
$\xi$ &-0.00 & 0.0036   &  0.0533 & -0.00   &0.0011   &-0.0030 & \\
$\phi$    &  1  &   0.0433  & 0.9688& 1  & 0.0145  & 1.0204  &   \\
$\tau_1$   &   -0.08 &  0.0001 &   -0.0801 & -0.08   &0.000059 &  -0.0802  &\\                                            
$\tau_2$    &   0.06  & 0.000103   &  0.0612& 0.06  &0.00003   &0.0595 & \\
$\sigma_u$  &  0.4   & 0.00139 &     0.3667& 0.4  & 0.00103  &  0.3681 &     \\     
\hline                                
5\%VaR      &-1.8547   &  0.0051   & -1.8738 & -1.8567  &0.0024 & -1.8684  &  \\
1\%VaR     & -2.6834 & 0.0089&   -2.6402 &  -2.6524 &0.00572 &-2.6433     &\\
5\%ES      & 0.1224 &  0.000022& 0.1236 &0.1225 & 0.00001 &0.1233 & \\
1\%ES     & 0.0310 &    0.000001   &  0.0306 &0.0307  & 0.0000007   & 0.0305      &     \\
\hline
\end{tabular}
 \end{table}

\begin{table}[t]
\caption{Summary statistics for the estimator of the RHYGARCH-tG model, data simulated from Model 2 }
\label{tab2}
\begin{tabular}{@{}llllllll} 
\hline
RHGARCH$\left(1,d,1\right)$ &   & T=1000 &  & &  T=3000  & &\\
par & True    &    MSE   &   Mean & True  & MSE &  Mean  &\\
\hline  
$\omega$ & 0.1 &  0.0022 & 0.0911 &   0.1   & 0.0020 &0.084 & \\
$\gamma$ & 0.1    & 0.0208 &0.1931   & 0.1   & 0.0110 & 0.1770    &\\
$\beta$  & 0.4   & 0.0244 &  0.3316& 0.4  & 0.0143 & 0.3592&  \\
$\delta$ & 0.4 & 0.0352 & 0.3204   &0.4 & 0.0216  &  0.32533       &\\
$d$      & 0.4 &  0.0167 &0.3779&0.4    & 0.0070 & 0.3762    &    \\
$\nu$      & 3&  0.0572   & 3.1787& 3     &0.0455 & 3.1712&    \\
$\xi$ &-0.00&   0.0066 & 0.069&  -0.00   & 0.0069 &0.068 & \\
$\phi$    &  1&  0.0138  &  0.9218& 1   & 0.0194  & 0.9304& \\
$\tau_1$   &   -0.08 &0.0022   & -0.0760 & -0.08 &0.00009 &-0.07602 &\\                                            
$\tau_2$    &   0.06  & 0.000038    &0.0566 & 0.06 & 0.000097  & 0.0569  & \\
$\sigma_u$  &  0.4   &  0.0011 &    0.3708 &0.4   & 0.0010  & 0.3725   &     \\  
\hline                                           
5\%VaR      & -2.625  & 0.0376  & -2.5672 & -2.668 & 0.0262  & -2.5652   &  \\
1\%VaR      &  -4.9864 &0.2118 &  -4.836& -4.988  &  0.1679 &  -4.8422 &\\
5\%ES      & 0.227   & 0.00043    & 0.2179 & 0.231 & 0.00034   & 0.2178 & \\
1\%ES     &0.0776   & 0.000082      & 0.0736& 0.0777    & 0.000067   & 0.0738    &     \\
\hline
\end{tabular}
\end{table}

\noindent Estimation results are summarized in  Tables \ref{tab1} and \ref{tab2}. Inspection of these tables reveals that, for all sample sizes, the QMLE procedure performs relatively well. Particularly, the MSE for the parameters $\epsilon$, $\tau_1$, $\tau_2$ and $\sigma_u$ are very small indicating that estimators are consistent. Moreover, we notice that the bias as well as the MSE decreases when the sample size increases. We observe also that parameter estimates are more precise in the measure equation than in the GARCH equation.\\ 
\noindent Another finding is that the ES has lowest bias estimation under $t$-student distribution when the sample size increases and inversely under the Gaussian distribution. In addition, the results for ES and VaR of the RHYGARCH-GG model deliver more satisfactory estimates than the RHYGARCH-tG model. Suggesting that the RHYGARCH-GG model can be used to forecast the  ES and VaR. The findings of this research are consistent with those from Gerlach and Chaowang \cite{gerlach2016forecasting}.

\section{Conclusion}
In this work,  the realized HYGARCH process is studied which  generalizes the realized GARCH model of Hansen et \textit{al.} in \cite{hansen2012realized} and the FLoGARCH model introduced by Vander Elst in \cite{vander2015fir}. Under some assumptions, the model shows to be strictly and weak stationary. The parameter estimation problem is addressed using the quasi-maximum likelihood procedure. Finite sample behaviors of this method were studied using Monte Carlo simulations. It indicates that the approach can yield asymptotic efficient estimates. The simulation shows that the RHYGARCH-tG model deliver more adequate estimates than the RHYGARCH-GG model for forecasting ES. Nevertheless, it shows that the  RHYGARCH-GG model has more precise estimates than the RHYGARCH-tG for forecasting the VaR. Since the results from the estimation methodology are encouraging, it will be interesting to examine, in a future work, the empirical application of the realized HYGARCH model in financial data.

\newpage
\begin{appendix}

\section{Appendix section}
\begin{proof}Lemma \ref{le1}
\noindent Denote by $v_t=\xi +\tau(z_t)+u_t$. By (\ref{eq6}) and (\ref{eq9}), we have

\begin{eqnarray*}\
\tilde{h}_t&=&\omega +\sum_{i_1=1}^{\infty}\psi_{i_1}\left(\phi \tilde{h}_{t-i_1}+v_{t-i_1} \right) \nonumber  \\  
&=&\omega +\sum_{i_1=1}^{\infty}\psi_{i_1}v_{t-i_1}+\phi\sum_{i_1=1}^{\infty}\psi_{i_1}\tilde{h}_{t-i_1}  \nonumber \\ 
&=&\omega +\sum_{i_1=1}^{\infty}\psi_{i_1}v_{t-i_1}+\phi\sum_{i_1=1}^{\infty}\psi_{i_1}\left[\omega +\sum_{i_2=1}^{\infty}\psi_{i_2}\left(\phi \tilde{h}_{t-i_1-i_2}+v_{t-i_1-i_2}\right)\right] \nonumber \\
&=&\omega+\sum_{i_1=1}^{\infty}\psi_{i_1}\left( v_{t-i_1}+ \phi\omega \right)+\phi\sum_{i_1,i_2=1}^{\infty} \psi_{i_1}\psi_{i_2}v_{t-i_1-i_2}+\phi^2\sum_{i_1,i_2=1}^{\infty}\psi_{i_1}\psi_{i_2}\tilde{h}_{t-i_1-i_2}\nonumber\\
&=& \omega+\sum_{i_1=1}^{\infty}\psi_{i_1}\left( v_{t-i_1}+ \phi\omega \right)+\phi\sum_{i_1,i_2=1}^{\infty} \psi_{i_1}\psi_{i_2} v_{t-i_1-i_2}
\nonumber \\
&+& \phi^2\sum_{i_1,i_2=1}^{\infty}\psi_{i_1}\psi_{i_2}(\omega+\sum_{i_3=1}^{\infty}\psi_{i_3}\tilde{x}_{t-i_1-i_2-i_3})\nonumber\\
&=&\omega+\sum_{i_1=1}^{\infty}\psi_{i_1}\left( v_{t-i_1}+ \phi\omega \right)+ \sum_{i_1,i_2=1}^{\infty} \phi\psi_{i_1}\psi_{i_2}\left(v_{t-i_1-i_2}
+\phi\omega\right)\nonumber \\
&+&\phi^2\sum_{i_1,i_2,i_3=1}^{\infty}\psi_{i_1}\psi_{i_2}\psi_{i_3}\tilde{x}_{t-i_1-i_2-i_3} \nonumber 
\end{eqnarray*}
for m step we have 
\begin{eqnarray*}
\tilde{h}_t&=&\sum_{l=0}^m \sum_{i_1,\cdots,i_l}^\infty \phi^{l-1}\psi_{i_1}\cdots\psi_{i_l}\left(\omega\phi+v_{t-i_1-\cdots-i_l}\right)\nonumber\\
 & & + \phi^{m}\sum_{i_1,\cdots,i_{m+1}}^\infty\psi_{i_1}\cdots\psi_{i_{m+1}}\tilde{x}_{t-i_1,\cdots,i_{m+1}}\\
 \tilde{h}_t&=&\sum_{l=0}^m \sum_{i_1,\cdots,i_l}^\infty \phi^{l-1}\psi_{i_1}\cdots\psi_{i_l}\left(\omega\phi+v_{t-i_1-\cdots-i_l}\right)\nonumber\\
 & & + (\sum_{i=1}^\infty\phi\psi_{i})^m \sum_{i_{m+1=1}}^{\infty}\psi_{i_{m+1}}\tilde{x}_{t-i_1,\cdots,i_{m+1}}. 
\end{eqnarray*}
If $m\rightarrow\infty$ and conditions (\ref{eq11}) are satisfied then

$\tilde{h}_t=\sum_{l=0}^{\infty}H_l\left(t\right).$
\end{proof}
\begin{proof} Theorem (\ref{theo1})
\begin{enumerate}
\item We show that condition (\ref{eq13}) implies the existence of a first moment of the bivariate process $\left(\tilde{h}_t,\tilde{x}_t\right)$ .\\\\
Taking the expectations in (\ref{eq6}) and (\ref{eq9}) and solving the linear system gives that if the bivariate process $\left(\tilde{h}_t,\tilde{x}_t\right)$ is mean-stationary, then 
\[
E\left(\tilde{h}_t\right)=\frac{\omega+\xi\sum_{i=1}^{\infty}\psi_i}{1-\phi\sum_{i=1}^{\infty}\psi_i}=\frac{\omega+\xi\psi(1)}{1-\phi\psi(1)}
\] and 
\[
E\left( \tilde{x}_t\right)=\frac{\xi+\phi\omega}{1-\phi\psi(1)}.
\]

\noindent Let's now proved the sufficient condition for the existence of the second moment of the process $\tilde{h}_t$. Applying the Minkowski inequality norm to (\ref{eq12}) in conditions (\ref{eq14}) , we get :

\[
E\left(\tilde{h}_t^2\right)^{\frac{1}{2}}\leq\sum_{l=0}^{\infty}\sum_{i_1,i_2,\cdots, i_l=1}^{\infty}\phi^{l-1}\psi_{i_1}\cdots\psi_{i_l}\left[E\left(\omega\phi +v_{t-i_1-i_2-\cdots-i_l}\right)^2\right]^\frac{1}{2}. 
\]

\noindent Let define and denote by $B$ the above equality: 
\[
B=\sum_{l=0}^{\infty}\sum_{i_1,i_2,\cdots,i_l=1}^{\infty}\phi^{l-1}\psi_{i_1}\cdots \psi{i_l}\left[\omega^2\phi^2+2\xi\omega\phi +E\left(v_{t-i_1-i_2-\cdots-i_l}\right)^2\right]^\frac{1}{2}.
\]

\noindent We have 

\begin{eqnarray*}
E(v_{t-i_1-i_2-\cdots-i_l})^2&=&E[\xi+\tau_1z_{t-i_1-i_2-\cdots-i_l}+\tau_2(z_{t-i_1-i_2-\cdots-i_l}^2-1)
\nonumber\\
&&+u_{t-i_1-i_2-\cdots-i_l}]^2,
\end{eqnarray*}

\noindent by developing this expression and using the fact that $u_t$ and $z_t$ are mutually independent we get:

\[
E\left(v_{t-i_1-i_2-\cdots-i_l}\right)^2=\xi^2+\tau_1^2+\sigma_u^2-\tau_2^2+2\tau_1\tau_2E\left(z_0^3\right)+\tau_2^2E\left(z_0^4\right),
\]

\noindent so we have:

\[
B=\left[k+2\tau_1\tau_2E\left(z_0^3\right)+\tau_2^2E\left(z_0^4\right)\right]^\frac{1}{2}\sum_{l=0}^{\infty}\left(\sum_{i=1}^{\infty}\psi_i\phi\right)^l,
\]

\noindent where $k=\omega+\frac{2\xi\omega}{\phi}+\xi^2\phi^{-2}+\tau_1^2\phi^{-2}+\sigma_u^2\phi^{-2}-\tau_2^2\phi^{-2}$. If the conditions (\ref{eq14}) and Lemma \ref{le1} are satisfied, the second moment of the process $\left(\tilde{h}_t\right)$ exists.\\
\end{enumerate}
\end{proof}
\begin{proof}Theorem \ref{theo2}
\begin{enumerate}
 \item Since it is very easy to verify that (\ref{eq12}) is a stationary solution, therefore we show that it is the unique strictly stationary solution. Assume that $y_t $ is any strictly stationary solution with finite first moment $E\left(y_0\right)$. Then, applying relations (\ref{eq9}) and (\ref{eq6}) after $m$ steps we obtain.
 
 \begin{eqnarray*} 
 y_t&=&\sum_{l=0}^m \sum_{i_1,\cdots,i_l}^\infty \phi^{l-1}\psi_{i_1}\cdots\psi_{i_l}\left(\omega\phi+v_{t-i_1-\cdots-i_l}\right)\nonumber\\
 & & + \sum_{i_1,\cdots,i_{m+1}}^\infty\phi^{m}\psi_{i_1}\cdots\psi_{i_{m+1}}\left(\phi y_{t-i_1-\cdots-i_{m+1}}+ v_{t-i_1-\cdots-i_{m+1}}\right).
  \end{eqnarray*}
  
 \noindent Therefore, we have 
 
 \begin{eqnarray}\label{eq15}
 y_t-h_t&=&\sum_{i_1,\cdots,i_{m+1}}^\infty\phi^{m}\psi_{i_1}\cdots\psi_{i_{m+1}}\left(\phi y_{t-i_1-\cdots-i_{m+1}}+ v_{t-i_1-\cdots-i_{m+1}} \right)\nonumber\\
 &&-\sum_{l=m+1}^\infty \sum_{i_1,\cdots,i_l}^\infty \phi^{l-1}\psi_{i_1}\cdots\psi_{i_l}\left(\omega\phi+v_{t-i_1-\cdots-i_l}\right).
\end{eqnarray}
\noindent Applying Chebychev's inequality to the first term on the right-hand side of (\ref{eq15}), we obtain 
\begin{eqnarray*}
\varepsilon P\left[\sum_{i_1,\cdots,i_{m+1}}^{\infty}\phi^{m}\prod_{i=i_1}^{i_{m+1}}\psi_{i}\left(\phi y_{t-\sum_{j=i_1}^{i_{m+1}}j}+ v_{t-\sum_{k=i_1}^{i_{m+1}}k}\right)>\varepsilon\right]&\leq&\left(E\left(y_0\right)+\frac{\xi}{\phi}\right){}
\nonumber\\
&\times&\left(\phi\sum_{i=1}^{\infty}\psi_i\right)^{m+1}.
\end{eqnarray*}
\noindent By (\ref{eq11}) and the Borel-Cantelli lemma, this implies almost sure convergence to zero as $m \rightarrow \infty$. We have $\sum_{l=0}^\infty H_l\left(t\right)< \infty$, choosing $m$ large enough, the second term on the right-hand side of (\ref{eq15}) can be made small with probability 1.Thus $h_t=y_t\; a.s$.
 \item According to condition (\ref{eq14}) and lemma \ref{le1}, the second moment of the process exists.To verify that the sequence $\tilde{h}_t $ defined by (\ref{eq12}) is weakly stationary, observe that:
 \[
E\left(\tilde{h}_t\right)=\frac{\omega+\xi\sum_{i=1}^{\infty}\psi_i}{1-\phi\sum_{i=1}^{\infty}\psi_i}=\frac{\omega+\xi\psi(1)}{1-\phi\psi(1)}
\]
  \begin{eqnarray*}
     	Cov\left(\tilde{h}_i,\tilde{h}_{i+t}\right)&=&E\left(\tilde{h}_i,\tilde{h}_{i+t}\right)-E\left(\tilde{h}_i\right)E\left(\tilde{h}_{i+t}\right)\nonumber\\
     	&=&\sum_{l,k=1}^\infty\sum_{i_1,i_2,\cdots i_l=1}^{\infty}\sum_{j_1,j_2,\cdots,j_k=1}^{\infty}\phi^{l+k-2} \psi_{i_1}\psi{i_2}\cdots\psi_{i_l}\psi_{j_1}\psi{j_2}\cdots\psi_{j_k}\nonumber\\
	&&Cov\left(v_{t-i_1-\cdots-i_l},v_{t-j_1-\cdots-j_k}\right)-\left(\frac{\omega+\xi\psi(1)}{1-\phi\psi(1)}\right)^2\nonumber\\
     	&=&Cov\left(\tilde{h}_0,\tilde{h}_t\right).
	\end{eqnarray*}	
\noindent Unicity results is obtained using the same lines as in 1.
\end{enumerate}
     \end{proof}
\end{appendix}

%Giraitis, L., Leipus, R., and Surgailis, D. (2009). ARCH(?) models and long memory. In T. G. Anderson, R. A. Davis, J. Kreis, and T. Mikosch (Eds.), Handbook of Financial Time Series (p. 71-84). Berlin: Springer Verlag.

\end{document}